\documentclass{appolb}
\usepackage{epsfig}

\usepackage{amssymb}

\begin{document}
\title{Heavy ions theory review
\thanks{Presented at Physics at LHC, Cracow (Poland) July 2006}%
}
\author{Carlos A. Salgado \footnote{Permanent address: Departamento de F\'\i sica de Part\'\i culas, Universidade de Santiago de Compostela (Spain).}
\address{Dipartimento di Fisica, Universit\`a  di Roma ``La Sapienza''\\ and INFN,
Roma, Italy}
}
\maketitle
\begin{abstract}
Some of the new developments in the theory of heavy ion collisions are reviewed. Much of the last progress have been triggered by the high energies available at RHIC. In the near future, the LHC will extend the energy reach in heavy ions by a factor thirty and give access to new QCD regimes characterized by large densities and temperatures and corresponding modified evolution equations.
\end{abstract}

\section{Introduction}

The advent of collider energies to heavy--ion physics is leading to a profound change  in the field. The new tools available, specially the access to the large transverse momentum part of the spectrum, allow for an unprecedented characterization of the high--density state created in such collisions. Accordingly, the traditional goal of producing a {\it Quark-Gluon Plasma}, the deconfined state of quarks and gluons predicted from QCD, has been enlarged to the study of several other mechanisms as thermalization, parton distribution functions at very small-$x$ or in--medium evolution of parton showers. In general words, a new line is emerging which attempt to study how the collective properties of the fundamental interactions appear. For that end, large energy densities in {\it extended} regions need to be produced in contrast with the traditional direction in high--energy physics which attempt to create the largest possible energy scales in well localized spacial regions for new physics to become observable.

The relevant questions which can be addressed in experiments of heavy ion collisions can be (artificially) classified depending on the time scale of the relevant phenomena: i) before the collision, the structure of the two Lorenz-contracted nuclei is rather different from a typical hadron at the same energy and collective phenomena, nowadays generically known under the name of Color Glass Condensate, appear; ii) once the high-density state is created, the relevant question is how thermalization and other collective mechanisms, as a possible hydrodynamical behavior, appear; iii) finally, the properties of the eventually equilibrated medium need to be studied by means of some indirect signals. The LHC is in an excellent position to address these questions \cite{exp}.

\subsection{Hard Probes provide a general framework}

Reaching collider energies provides the additional tool of the hard part of the spectrum, characterized by large virtualities, to be used for these studies. A typical hard cross section can be written in the form
\begin{equation}
\sigma^{AB\to h}=
f_A(x_1,Q^2)\otimes f_B(x_2,Q^2)\otimes \sigma(x_1,x_2,Q^2)\otimes D_{i\to h}
(z,Q^2)\, ,
\label{eqhard}
\end{equation}
with a factorization between the short-distance perturbative cross section $\sigma(x_1,x_2,Q^2)$, computable in powers of $\alpha_s(Q^2)$ and two long-distance terms, the proton/nuclear parton distribution functions (PDF), $f_A(x,Q^2)$, encoding the partonic structure of the colliding objects, and the fragmentation functions (FF), $D(z,Q^2)$, describing the hadronization of the parton $i$ into a final hadron $h$. These long-distance terms are non-perturbative objects, but whose evolution can be computed in perturbative QCD. These are precisely the objects which will be modified in the case that the {\it extension} of the colliding system interferes with the dynamics, while the short-distance part is expected to remain unchanged if the virtuality is large enough. This interference between geometry and dynamics makes hard probes perfect tools to characterize the medium properties through the modification of the long-distance terms in (\ref{eqhard}).

A conceptually simple example is the case of the $J/\Psi$ whose production cross section is, schematically,   
\begin{equation}
\sigma^{hh\to J/\Psi}=
 f_i(x_1,Q^2)\otimes f_j(x_2,Q^2)\otimes
\sigma^{ij\to [c\bar c]}(x_1,x_2,Q^2)
 \langle {\cal O}([c\bar c]\to J/\Psi)\rangle\, ,
\end{equation}
where now $ \langle {\cal O}([c\bar c]\to J/\Psi)\rangle$ describes the hadronization of a $c\bar c$ pair in a given state (for example a color octet) into a final $J/\Psi$. This long-distance part is expected to be modified in a medium at finite temperature in which the deconfinement avoids the bound states to exist \cite{Matsui:1986dk}. However, this modification, being non-perturbative, lacks of good theoretical control and the real situation about the $J/\Psi$-suppression is complicated \cite{tram}. 

From the computational point of view, a theoretically simpler case involves the modification of the {\it evolution} of both the parton distribution and the fragmentation functions in a dense or finite--temperature medium. This needs of large scales $Q^2$ in order for the strong coupling to be small and perturbative methods to apply. In general, the presence of a dense medium is translated into non-linear terms in the evolution equations. 

\section{The initial state: the Color Glass Condensate}

\begin{figure}
\begin{minipage}{0.35\textwidth}
\begin{center}
\includegraphics[width=\textwidth]{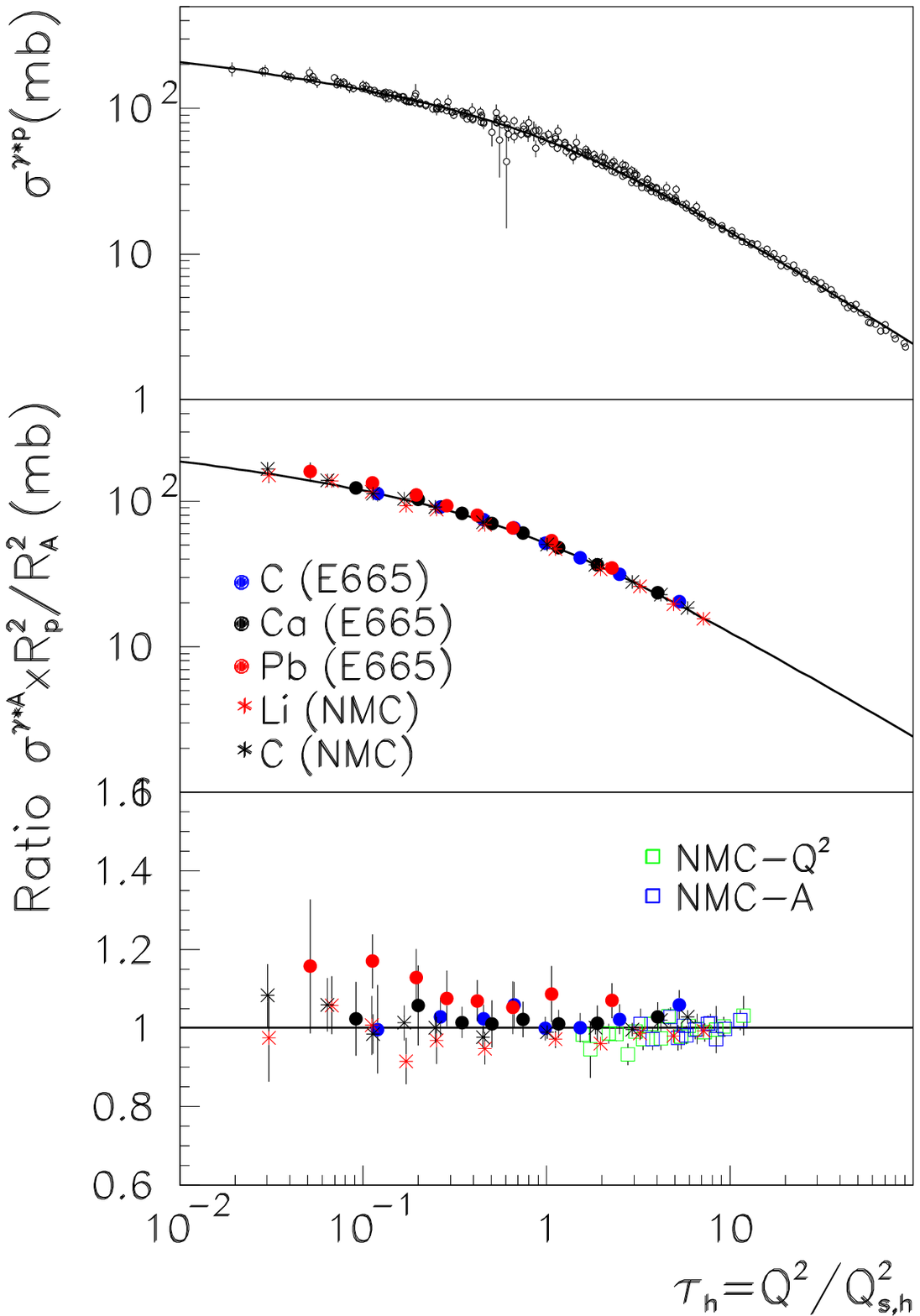}
\end{center}
\end{minipage}
\hfill
\begin{minipage}{0.65\textwidth}
\begin{center}
\includegraphics[width=0.8\textwidth]{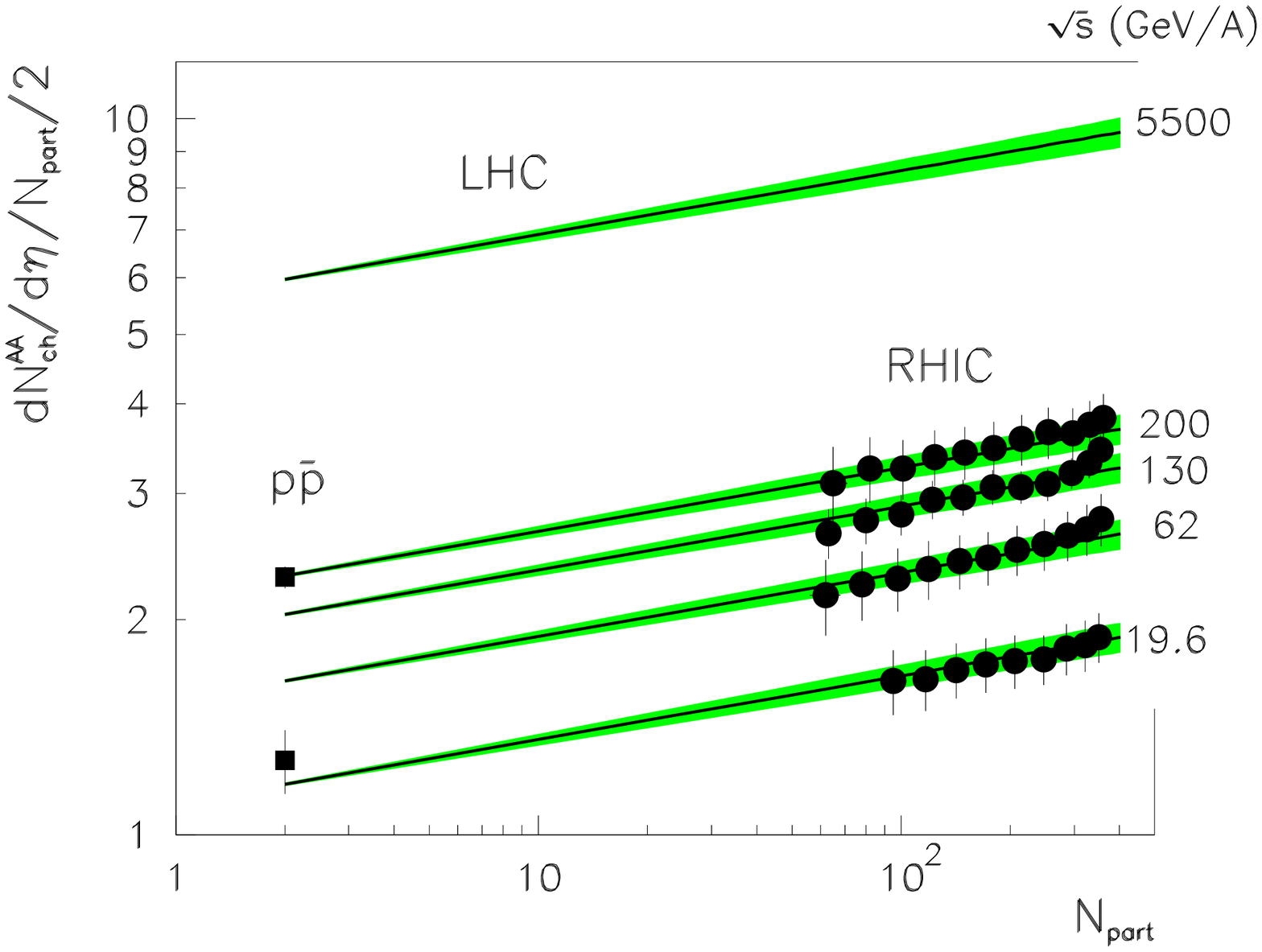}
\end{center}
\end{minipage}
\caption{Left: Geometric scaling in lepton-proton \protect\cite{Stasto:2000er} and lepton-nucleus \protect\cite{Armesto:2004ud} data. Right: Central rapidity multiplicities in $p\bar p$ and AuAu collisions at different centralities \protect\cite{Back:2004je} and the corresponding description from Eq. (\ref{eqmult})\protect\cite{Armesto:2004ud}.}
\label{fig1}
\end{figure}

The study of the modifications of the nuclear partonic distributions is nowadays known under the generic name of {\it Color Glass Condensate}. In its original formulation \cite{McLerran:1993ni} it provides a general framework for the whole collision, based on an effective theory separating the fast modes in the nuclear wave function from the {\it generated} slow modes, associated to small-$x$ gluons. This small-$x$ gluons have parametrically large (${\cal O}(1/\alpha_s)$) densities and can be treated as classical fields. The quantum evolution equation of this setup is known and, remarkably, can be written in a rather simple form in the large-$N_C$ limit \cite{Iancu:2003xm}. A sophisticated technology has been developed in the last decade in this framework which, in addition to describe the structure of the incoming wave functions, aims to provide the link to the subsequent evolution into a thermal system \cite{Lappi:2006fp}. From a phenomenological point of view, the most successful applications of these formalism are the description of the multiplicities measured at RHIC \cite{Kharzeev:2000ph, Armesto:2004ud}; the possible presence of the predicted geometric scaling in lepton-hadron data \cite{Stasto:2000er,Freund:2002ux,Armesto:2004ud}; and the suppression of inclusive particles at forward rapidities at RHIC \cite{Arsene:2004ux}, which has been predicted as a result of small-$x$ quantum evolution \cite{forwardsup}. A particularly economic description is presented in Fig. \ref{fig1}. Here, the saturation scale is obtained from lepton-proton and lepton-nucleus data as $Q^2_{\rm sat}\propto x^{-\lambda}A^{1/3\delta}$ with fitted parameters $\lambda=0.288$ and $\delta=0.79$. Fig. \ref{fig1} left shows the quality of the geometric scaling in lepton-proton \cite{Stasto:2000er} and lepton-nucleus \cite{Armesto:2004ud} data. Assuming the same scaling to hold in AA collisions, the multiplicity in the central rapidity can be written as \cite{Armesto:2004ud}
\begin{equation}
\frac{1}{N_{\rm part}}
\frac{dN^{AA}}{d\eta}\Bigg\vert_{\eta\sim 0}=N_0\sqrt{s}^\lambda
N_{\rm part}^{\frac{1-\delta}{3\delta}}\, .
\label{eqmult}
\end{equation}
where only a total normalization factor $N_0$ is needed once the energy and centrality dependences are fixed by lepton-proton and lepton-nucleus data respectively. Fig. \ref{fig1} shows the comparison of this simple formula with available data \cite{Back:2004je}. 

\section{The medium--modification of jet properties}

\label{sec:jq}

A perturbative modification of the fragmentation of the highly virtual produced partons, has been proposed as a tool to characterize the medium produced in heavy ion collisions \cite{Salgado:2005pr}. These effects on high-$p_t$ particles, generically known as {\it jet quenching}, constitute one of the main experimental observations at RHIC so far \cite{Adcox:2001jp}. In the near future, the LHC will extend the range in transverse momentum for at least one order of magnitude \cite{exp}. In the vacuum, a perturbatively produced high-$p_t$ parton with virtuality $Q^2\sim p_t^2$ develops a parton shower by radiating other partons (mainly gluons) with decreasing virtuality. This shower stops when a typical hadronic scale ${\cal O}(1 $GeV$^2)$ is reached. The resulting jet is an extended object in which the collinear and soft emitted gluons are emitted inside a cone around the original parton direction. When the high-$p_t$ particle is produced in the medium created after a heavy ion collision, this parton shower is modified. The jets, being extended structures, provide an excellent tool to characterize the medium properties at different scales. 

At high enough parton energies, the main mechanism driving the modifications of high-$p_t$ evolution is the medium-induced gluon radiation. As in general high-energy processes, the propagation of the partons through the medium can be described in terms of Wilson lines averaged in the allowed configuration of a medium. Several prescriptions exist for these averages and in the multiple soft scattering limit, the saddle point approximation of the Wilson lines define a single parameter of the medium, the transport coefficient, $\hat q$, given by the average transverse momentum squared acquired by the gluon per mean free path. The typical energy spectrum of radiated gluons is softer than in the vacuum although formation time effects provide a typical scale $\hat \omega\sim \hat q^{1/3}$ under which the radiation is suppressed. In the same way, the angular distribution is regulated at angles smaller than $\sin\hat\theta\simeq(\hat q/\omega^3)^{1/4}$. As a result of formation time effects (or coherence effects) the medium-induced gluon radiation is, hence, finite in the infrared or collinear limits. The two main predictions are the suppression of high-$p_t$ yields due to additional in-medium energy loss and the associated broadening of the jet structures.

\subsection{The suppression of high-$p_t$ particles}

\begin{figure}
\begin{center}
\includegraphics[width=0.43\textwidth,angle=-90]{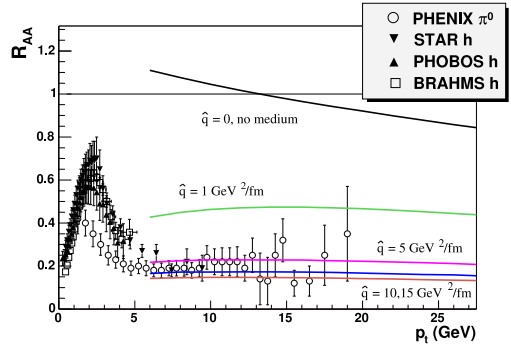}
\end{center}
\caption{Nuclear modification factor, $R_{AA}$, in central
AuAu collisions at $\sqrt{s}$=200 GeV \protect\cite{Eskola:2004cr}. Data from \protect\cite{Adcox:2001jp}.}
\label{figraa}
\end{figure}

Although the degree of theoretical refinement of the jet quenching formalisms is still not completely satisfactory, it provides a good description of experimental data with the medium-modified fragmentation function computed as
\begin{equation}
D_{i\to h}^{\rm med}(z,Q^2)=P_E(\epsilon)\otimes D_{i\to h}(z,Q^2)
\label{eqff}
\end{equation}
here, $P_E(\epsilon)$ gives the probability of an additional in-medium energy loss and is normally computed by assuming a simple Poisson distribution with the medium-induced gluon radiation as input \cite{Baier:2001yt,Salgado:2003gb}. 
Once the geometry of the system is correctly taken into account, a fit to RHIC data (see Fig. \ref{figraa}) gives the value of the time-averaged transport coefficient $\hat q\sim 5...15$ GeV$^2$/fm. This large value and the corresponding uncertainty is a direct consequence of the surface trigger bias effect in inclusive particle suppression measurements \cite{Muller:2002fa,Eskola:2004cr,Dainese:2004te}. This is an intrinsic limitation of inclusive measurements on the characterization of the medium and on the study of the dynamics underlying the propagation of highly energetic partons through a dense medium. Further constraints can be found by i) measuring different particles species, and in particular heavy quarks, as the formalism predicts the hierarchy $\Delta E_g>\Delta E_q^{\rm m=0}>\Delta E_Q^{\rm m\neq 0}$; ii) by directly measuring the induced radiation, i.e. by reconstructing the jet structure in a heavy ion collision.

\begin{figure}
\begin{center}
\includegraphics[width=0.36\textwidth,angle=-90]{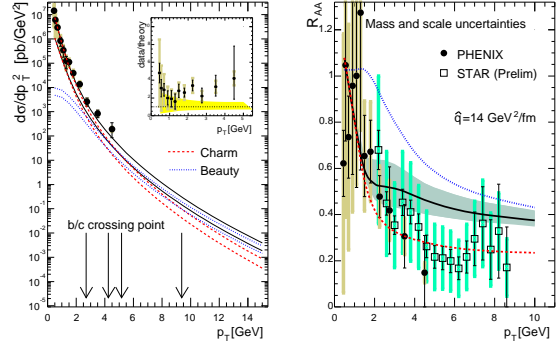}
\end{center}
\caption{Left: Comparison of the FONLL calculation of single 
inclusive electrons from pp collisions at $\sqrt{s}$=200GeV \protect\cite{Cacciari:2005rk}. Right: The nuclear modification factor of electrons with the corresponding uncertainty coming from the perturbative benchmark on the relative $b/c$ contribution. Figure from \protect\cite{Armesto:2005mz}; data from \protect\cite{elqm05}.}
\label{fighq}
\end{figure}

New data from RHIC on non-photonic electrons \cite{elqm05} attempt to answer the question on heavy quark in-medium energy loss. These electrons are expected to come from the decays of charm and beauty quarks. The perturbative description of the relative contribution of both quarks to the electron yield is, however, not under good control and the $b/c$ crossing point can be as low as 2 GeV or as large as 10 GeV when the usual mass and scale uncertainties are taken into account in the   approximation. This translates into a large uncertainty in the ratio $R_{AA}^e$ as can be seen in Fig. \ref{fighq} \cite{Armesto:2005mz}. The description of the experimental data is reasonable within the error bars, although not completely satisfactory. A clear distinction between heavy mass effects in medium-modified gluon radiation seems only possible with a better identification of the $c$ and $b$ contributions and, ideally, by a direct measurement of the $D$ and $B$ mesons. 

\subsection{Jet shapes in opaque media}

The modification of the jet properties is thought to be the most powerful tool to investigate the medium properties as well as the dynamics underlying jet-quenching.  From an experimental point of view, the callibration uncertainty of true jets in the large background environment of a heavy--ion collision is the main issue to overcome \cite{exp}. The identification of observables with small sensitivity to background subtractions is of primary importance \cite{Salgado:2003rv}. From a theoretical point of view, studying medium--modification of jet shapes in the most general case is complicated and likely would need a full Monte Carlo implementation of a in-medium parton shower process, which is not yet available. Here, we quote a simplified study \cite{Polosa:2006hb} which provides a quite good phenomenological understanding of the surprising non-gaussian shape found in two particle correlations at RHIC \cite{Grau:2005sm}. The main observation is that for energies smaller than $\hat\omega\simeq 2(2\hat q)^{1/3}$ the typical medium-induced radiation angle reaches its maximum value $\sin\theta=1$. In this conditions, the medium-induced gluon radiation spectrum has the simple form \cite{Polosa:2006hb}
\begin{equation}
\frac{dI^{\rm med}}{d\omega\,dk^2_\perp}\simeq\frac{\alpha_s C_R}{16\pi}\,L\,\frac{1}{\omega^2}.
\label{eq:cohspec}
\end{equation}
Taking a typical value $\hat q=10$ GeV$^2$/fm, Eq. (\ref{eq:cohspec}) is valid for 
$\omega<\hat\omega\simeq$ 3 GeV.
Compared to the corresponding spectrum in the vacuum, 
\begin{equation}
\frac{dI^{\rm vac}}{dz dk_\perp^2}=\frac{\alpha_s}{2\pi}\frac{1}{k_\perp^2}P(z),
\label{eq:specvac}
\end{equation}
the one in
(\ref{eq:cohspec}) is softer and the typical emission angles larger, producing a broadening of the jet signals. 

For practical applications, exclusive distributions, giving the probability of one, two... emissions are needed. How to construct such probabilities, using Sudakov form factors 
\begin{equation}
\Delta(t)\equiv\exp\left [-\int_{t_0}^t\frac{dt'}{t'}\int dz
\frac{\alpha_S}{2\pi}P(z)\right],
\label{eqsudak}
\end{equation}
is a well known procedure in the vacuum. In \cite{Polosa:2006hb} it is proposed to generalize this procedure by defining eq. (\ref{eqsudak}) for the medium by just changing $dI^{\rm vac}/dzdk^2_\perp$ to $dI^{\rm med}/dzdk^2_\perp$ given by Eq.~(\ref{eq:cohspec}). There are convincing motivations to 
make this {\it Ansatz}: (i) it provides a clear probabilistic interpretation with the right limit to the most used ``quenching weights'' \cite{Baier:2001yt,Salgado:2003gb} when the virtuality is ignored; (ii) the evolution equations in the case of nuclear fragmentation functions in DIS can be written as usual DGLAP equations with medium-modified splitting functions $P(z)\to P(z)+\Delta P(z)$ \cite{Wang:2001if}. In our case Eq.~(\ref{eq:cohspec}) would correspond to $\Delta P(z)$ with the appropriate factors.

Two extreme cases were studied in \cite{Polosa:2006hb}: (1) one of the particles takes most of the incoming energy $\omega_1\gg\omega_2$ and, hence, $\theta_1\simeq 0$ -- ``J-configuration''; (2) the two particles share equally the available energy, $\omega_1\simeq\omega_2$ and $\theta_1\simeq\theta_2$ -- ``Y-configuration''.

{\it Case (1)}. The splitting probability in the laboratory variables $\Phi$ and $\eta$
\begin{equation}
\frac{d{\cal P}(\Phi,z)}{dz\,d\Phi}\Bigg\vert_{\eta=0}=
\frac{\alpha_s C_R}{16\pi^2}\,E\,L\,\cos\Phi
\exp\left\{-E\,L\,\frac{\alpha_s C_R}{16\pi}\cos^2\Phi\right\}
\label{eq:splitlab}
\end{equation}
presents a non-trivial angular dependence, which we stress has been found by the same perturbative mechanism and parameters describing the inclusive particle suppression. The distribution found has two maxima whose positions are determined by:
\begin{equation}
\Phi_{\rm max}=\pm {\rm arccos}\sqrt{\frac{8\pi}{E\,L\,\alpha_s\, C_R}}
\label{eq:phimax}
\end{equation}
\begin{figure}
\begin{minipage}{0.48\textwidth}
\begin{center}
\includegraphics[width=0.75\textwidth,angle=-90]{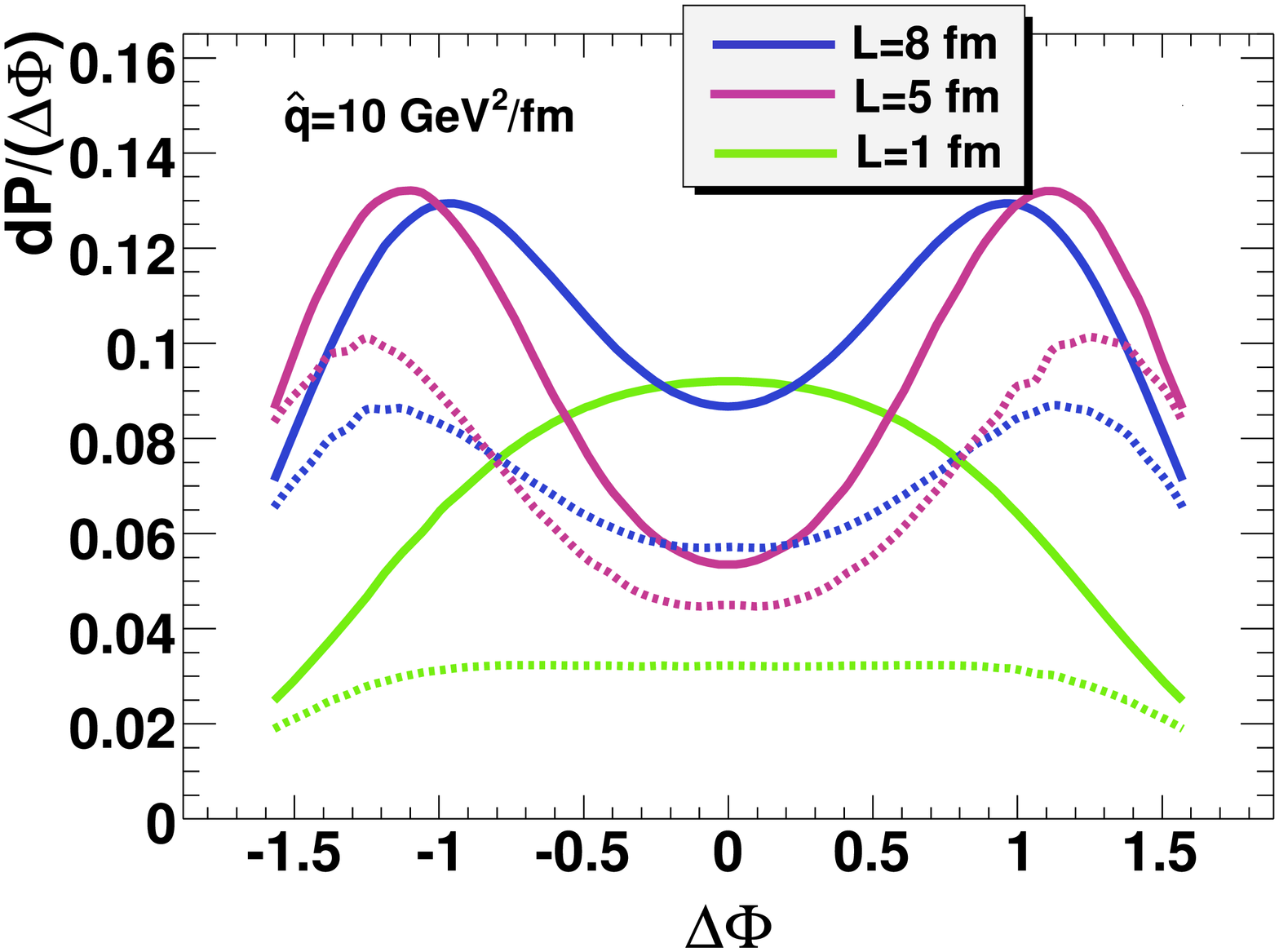}
\end{center}
\end{minipage}
\hfill
\begin{minipage}{0.48\textwidth}
\begin{center}
\includegraphics[width=0.75\textwidth,angle=-90]{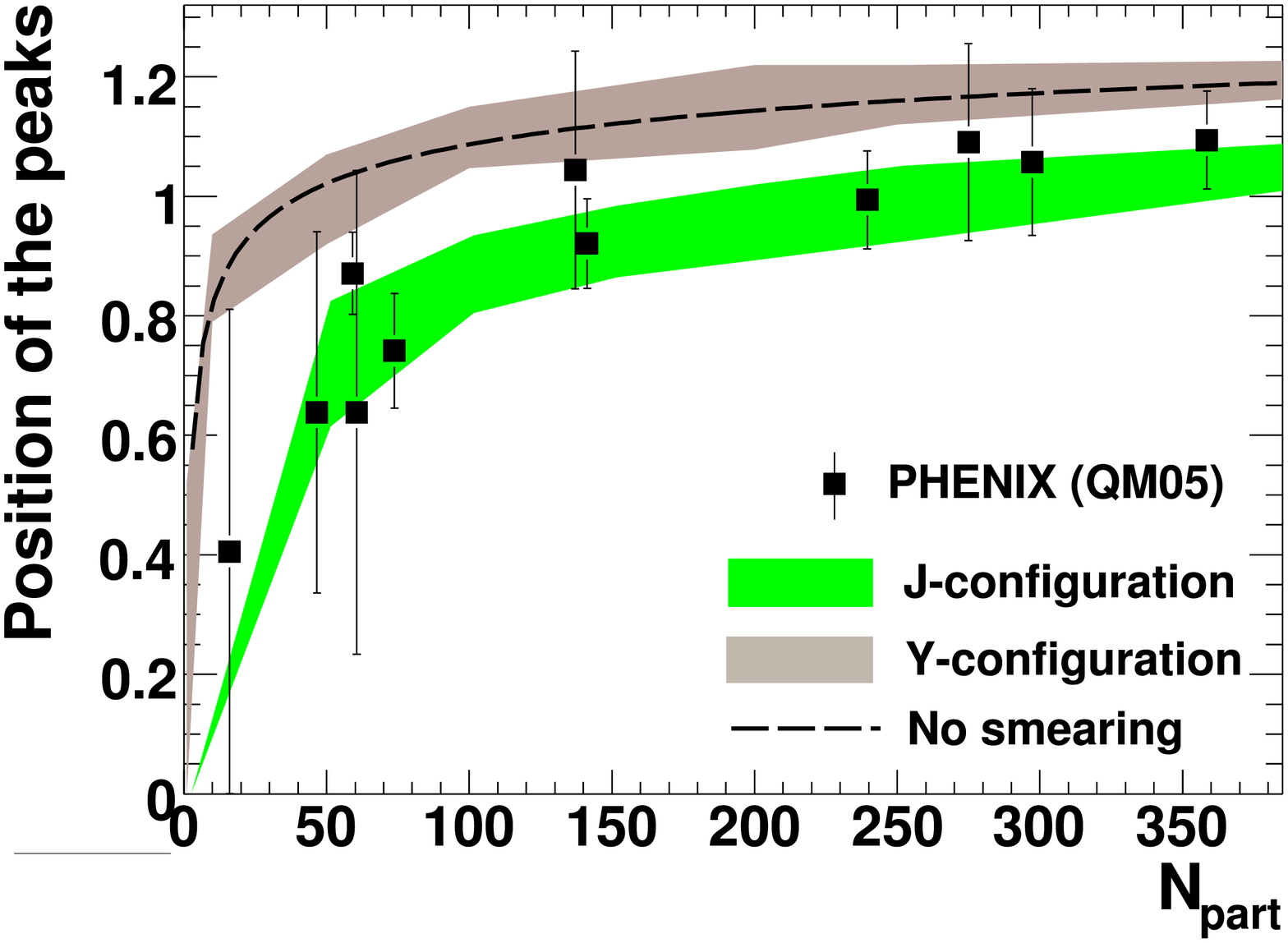}
\end{center}
\end{minipage}
\caption{Left: The probability of just one splitting (\protect\ref{eq:splitlab}) \cite{Polosa:2006hb} as a function of the laboratory azimuthal angle $\Delta\Phi$ for a gluon jet of $E_{\rm  jet}=7$ GeV. Different medium lengths are plotted for the  J- and Y-configurations (solid and dotted lines respectively). Right: Position of the peaks of the $\Delta\Phi$-distribution and comparison with PHENIX data from Ref. \cite{Grau:2005sm}.}
\label{fig:ndist}
\end{figure}

%
%

The angular shape in (\ref{eq:splitlab}) is very similar to the one found experimentally. In \cite{Polosa:2006hb} a simple model is proposed to take into account the different smearing effects from the experimental triggering conditions. The results are plotted in Fig. \ref{fig:ndist} for three different medium lengths and $E_{\rm jet}=7$ GeV. To compare with the measured centrality dependence on the position of the maxima \cite{Grau:2005sm} $L=N_{\rm part}^{1/3}$ was taken and plotted in Fig. \ref{fig:ndist} -- see \cite{Polosa:2006hb} for details.

The above explanation for the non-gaussian shapes found experimentally at RHIC has the appeal of presenting a unified description of all high-$p_t$ effects in heavy--ion collisions measured at RHIC. This effects could be further amplified for jets developing in a flowing medium \cite{Armesto:2004pt}. However, other mechanisms lead to similar away--side correlations. In the most popular one, these angular structures correspond to the shock waves produced by the highly energetic particle traversing the medium \cite{conical}. In this picture, a large amount of the energy lost must be transferred to the medium almost instantaneously and take part in the hydrodynamical evolution. In general the released energy excites both sound and dispersive modes, and only the first ones produce the desired cone--signal (in the dispersive mode, the energy travels basically collinear with the jet). The energy deposition needed for the sound modes to become visible in the spectrum has been found to be quite large \cite{Casalderrey-Solana:2006sq}. To our knowledge, no attempt has been made so far to describe the centrality dependence of the shape of the azimuthal correlations in this approach. 
Given the fact that the two formalisms rely on completely different hypotheses, 
finding experimental observables which could distinguish between them is certainly an issue which deserves further investigation. 

\section{Hydrodynamics and intermediate-$p_t$}

Together with the high-$p_t$, the most relevant data from RHIC so far is the measurement of azimuthal anisotropies in the momentum distributions, which is considered as the main check of themalization in the medium. As originally proposed \cite{Ollitrault:1992bk} the hydrodynamical evolution of initial spacial anisotropies lead to momentum anisotropies due to the presence of pressure gradients. Although not perfectly, the experimental data from RHIC at low and moderate-$p_t$ can be consistently described by considering the formation of a thermalized medium at the very early stages of the collision $\tau\lesssim 1$ fm, which evolves as a perfect fluid \cite{Eskola:2005ue} -- i.e. follows a hydrodynamical evolution with negligible viscosity. This observation has lead to the developing paradigm of an ideal fluid being created at RHIC \cite{Shuryak:2005pp}. This state would be a strongly coupled Quark Gluon Plasma, with properties far from those of the asymptotic case of a gas of free quarks and gluons. Whether this interpretation is correct or not is a matter of debate at present.

\subsection{Counting the valence quarks of exotic hadrons}

Very interesting effects appear in the intermediate region of $2\lesssim p_\perp\lesssim 6$ GeV/c. The most spectacular of them is the appearance of valence quark number scaling laws for baryons and mesons: (i) $R_{CP}$ seems to depend only on the valence number of the produced particle; (ii) the elliptic flow parameter $v_2$ is universal when plotted as $v_2(p_\perp/n)/n$, $n$ being the number of valence quarks. The most successful model to describe these features is a two component soft+hard model. In this model, the soft spectrum is supposed to come from the recombination of quarks in a medium in thermal equilibrium \cite{Friesvb}. Thus for a particle with $N$ valence quarks
\begin{equation}
  \frac{dN_{N}}{d^2 P_\perp dy}\Big|_{y=0} = 
  C_{N} M_\perp \frac{\tau A_\perp}{(2\pi)^3} \, 2\,
  \left(\prod_{k=1}^{N}\gamma_k\right) \,
  I_0 \left[ \frac{P_\perp \sinh \eta_\perp}{T}\right] k_N(P_\perp)\;.
  \label{eq:4qspec}
\end{equation}
$A_\perp=\pi \rho_0^2$ is the transverse area of the partonic medium at freeze-out, $\tau$ the hadronization time, $M_\perp$ is the transverse mass of the hadron, $C_N$ is the hadron degeneracy factor and $\gamma_k$ is the fugacity factor for a given 
quark $k$. 
The corresponding parameters can be found in \cite{Maiani:2006ia}. In \ref{eq:4qspec} the short-hand notation for a N-quark hadron
\begin{eqnarray}
  k_N(P_\perp) &=&  
  K_1 \left[ \frac{\cosh \eta_\perp}{T}\sum_{a=1}^N\sqrt{m_a^2 + \frac{P_\perp^2}{N^2}}\right] \, 
\label{eq:k4q}
\end{eqnarray}
has been introduced with $I_0$ and $K_1$ the corresponding Bessel functions.

\begin{figure}[htb]
\begin{center}
\includegraphics[width=0.55\textwidth]{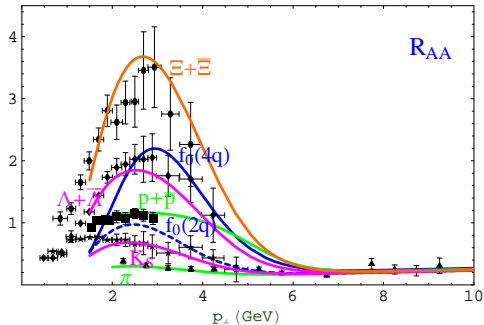}
\caption{
Starting from below: $R_{AA}$ for $\pi$, $K_S$, $f_0(980)$ as a $s\bar s$ state, $p+\bar p$, $\Lambda+\bar \Lambda$, $f_0(980)$ as a 4-quark state and $\Xi^-+{\bar \Xi}^+$. Data from Ref. \cite{Salur:2005nj}.}
\label{RAA:RHIC}
\end{center}
\end{figure}

The hard part of the spectrum is basically given by Eq. (\ref{eqhard}) with a simplified energy loss formalism, which constitutes an effective way of parameterizing the more refined one given in Section \ref{sec:jq}.
In \cite{Maiani:2006ia} this model has been extended to the case of a 4-quark meson to study the sensitivity of these observables to make a case for the discovery of exotic states in heavy-ion collisions, in particular for the $f_0(980)$. In Fig. \ref{RAA:RHIC} $R_{AA}$ for different mesons and baryons are plotted and compared with experimental data from RHIC. The conclusion from this figure is that if recombination is the right mechanism underlying the baryon/meson difference at intermediate $p_\perp$, heavy-ion collisions are ideal tools to study the quark content of different resonances and to find a definitive answer to the structure of these exotic states.

\section{New ideas}

The field of ultrarelativistic heavy--ion collisions has been, arguably, the most active field of research and discoveries in QCD in the last years. It is not unexpected that this will continue in the future, in particular within the unexplored regions accessible at the LHC. This activity is also reflected in the interest shown from other fields, in particular in the relation of experimental observables with string theory computable quantities by means of the AdS/CFT correspondence. The literature is already very vast, and impossible to quote here all the most relevant contributions; let just us mention the calculation of the shear viscosity \cite{Policastro:2001yc}; the heavy quark diffusion coefficient \cite{Casalderrey-Solana:2006rq}; the jet quenching transport coefficient \cite{Liu:2006ug}; the presence of shock waves \cite{Friess:2006fk}; the relation with hydrodynamics \cite{Janik:2005zt}.

\section*{Acknowledgements}
CAS is supported by the 6th Framework Programme of the European Community under the Marie Curie contract MEIF-CT-2005-024624.


\begin{thebibliography}{99}

\bibitem{exp}
See the contributions to these proceedings by G. Roland, E. Nappi, D. Hoffman, B. Wosiek, C. Jorgensen, B. Wyslouch, K. Oyama, R. Turrisi and A. Mastroserio.

\bibitem{Matsui:1986dk}
  T.~Matsui and H.~Satz,
  Phys.\ Lett.\ B {\bf 178} (1986) 416.

\bibitem{tram} 
  V.~N.~Tram  [PHENIX Collaboration],
  nucl-ex/0606017.

\bibitem{McLerran:1993ni}
  L.~D.~McLerran and R.~Venugopalan,
  Phys.\ Rev.\ D {\bf 49} (1994) 2233;
  Phys.\ Rev.\ D {\bf 49} (1994) 3352;
  Phys.\ Rev.\ D {\bf 50} (1994) 2225.


\bibitem{Iancu:2003xm}
 For recent reviews see: 
  E.~Iancu and R.~Venugopalan,
 hep-ph/0303204;
  D.~N.~Triantafyllopoulos,
  Acta Phys.\ Polon.\ B {\bf 36}, 3593 (2005);
  A.~Kovner,
  Acta Phys.\ Polon.\ B {\bf 36} (2005) 3551.


\bibitem{Lappi:2006fp}
  R.~Baier, A.~H.~Mueller, D.~Schiff and D.~T.~Son,
  Phys.\ Lett.\ B {\bf 502} (2001) 51;
  T.~Lappi and L.~McLerran,
  hep-ph/0602189;
  P.~Romatschke and R.~Venugopalan,
  hep-ph/0605045.

\bibitem{Kharzeev:2000ph}
  D.~Kharzeev and M.~Nardi,
  Phys.\ Lett.\ B {\bf 507} (2001) 121.


\bibitem{Armesto:2004ud}
  N.~Armesto, C.~A.~Salgado and U.~A.~Wiedemann,
  Phys.\ Rev.\ Lett.\  {\bf 94} (2005) 022002

\bibitem{Stasto:2000er}
  A.~M.~Stasto, K.~Golec-Biernat and J.~Kwiecinski,
  Phys.\ Rev.\ Lett.\  {\bf 86} (2001) 596; A.~M.~Stasto these proceedings.

\bibitem{Freund:2002ux}
  A.~Freund {\it et al.}
  Phys.\ Rev.\ Lett.\  {\bf 90} (2003) 222002 

\bibitem{Arsene:2004ux}
  I.~Arsene {\it et al.}  [BRAHMS Collaboration],
  Phys.\ Rev.\ Lett.\  {\bf 93} (2004) 242303.

\bibitem{forwardsup}
  D.~Kharzeev, E.~Levin and L.~McLerran,
  Phys.\ Lett.\ B {\bf 561} (2003) 93;
  R.~Baier, A.~Kovner and U.~A.~Wiedemann,
  Phys.\ Rev.\ D {\bf 68}, 054009 (2003);
  D.~Kharzeev, Y.~V.~Kovchegov and K.~Tuchin,
  Phys.\ Rev.\ D {\bf 68} (2003) 094013;
  J.~L.~Albacete {\it et al.}
  Phys.\ Rev.\ Lett.\  {\bf 92}, 082001 (2004).

\bibitem{Back:2004je}
  B.~B.~Back {\it et al.},
  Nucl.\ Phys.\ A {\bf 757} (2005) 28.

\bibitem{Salgado:2005pr}
  For a recent review see e.g. C.~A.~Salgado,
 hep-ph/0510062 and references therein.

\bibitem{Baier:2001yt}
  R.~Baier, Y.L.~Dokshitzer, A.H.~Mueller and D.~Schiff,
  JHEP {\bf 0109} (2001) 033

\bibitem{Salgado:2003gb}
  C.~A.~Salgado and U.~A.~Wiedemann,
  Phys.\ Rev.\ D {\bf 68}, 014008 (2003).

\bibitem{Muller:2002fa}
  B.~Muller,
  Phys.\ Rev.\ C {\bf 67} (2003) 061901;
%

\bibitem{Eskola:2004cr}
  K.~J.~Eskola {\it et al.}
  Nucl.\ Phys.\ A {\bf 747} (2005) 511.

\bibitem{Dainese:2004te}
  A.~Dainese, C.~Loizides and G.~Paic,
  Eur.\ Phys.\ J.\ C {\bf 38} (2005) 461;
  hep-ph/0511045.


\bibitem{Adcox:2001jp}
%
S.~S.~Adler {\it et al.}  [PHENIX Collaboration],
Phys.\ Rev.\ C {\bf 69} (2004) 034910;
%
%
J.~Adams {\it et al.}  [STAR Collaboration],
Phys.\ Rev.\ Lett.\  {\bf 91} (2003) 172302;
%
B.~B.~Back {\it et al.}  [PHOBOS Collaboration],
Phys.\ Lett.\ B {\bf 578} (2004) 297;
%
I.~Arsene {\it et al.}  [BRAHMS Collaboration],
Phys.\ Rev.\ Lett.\  {\bf 91} (2003) 072305.
%
  M.~Shimomura  [PHENIX],
  nucl-ex/0510023.

\bibitem{Cacciari:2005rk}
  M.~Cacciari, P.~Nason and R.~Vogt,
  Phys.\ Rev.\ Lett.\  {\bf 95} (2005) 122001

\bibitem{Armesto:2005mz}
  N.~Armesto, {\it et al.}
  Phys.\ Lett.\ B {\bf 637} (2006) 362.

\bibitem{elqm05}
S.S. Adler {\it et al.} [PHENIX] nucl-ex/0510047;
J. Bielcik [STAR] nucl-ex/0511005.

\bibitem{Salgado:2003rv}
  C.~A.~Salgado and U.~A.~Wiedemann,
  Phys.\ Rev.\ Lett.\  {\bf 93} (2004) 042301

\bibitem{Adler:2005ee}
  S.~S.~Adler {\it et al.}  [PHENIX Collaboration],
  nucl-ex/0507004.

\bibitem{Armesto:2004pt}
  N.~Armesto, C.~A.~Salgado and U.~A.~Wiedemann,
  Phys.\ Rev.\ Lett.\  {\bf 93} (2004) 242301;
%
  Phys.\ Rev.\ C {\bf 72} (2005) 064910;

\bibitem{conical}
 H.~Stoecker,
  Nucl.\ Phys.\ A {\bf 750}, 121 (2005);
J.~Casalderrey-Solana, E.~V.~Shuryak and D.~Teaney,
  hep-ph/0411315;
J. Ruppert and B.~Muller,
Phys.\ Lett.\ B {\bf 618} (2005) 123;

\bibitem{Polosa:2006hb}
  A.~D.~Polosa and C.~A.~Salgado,
  hep-ph/0607295.

\bibitem{Wang:2001if}
X.~N.~Wang and X.~F.~Guo,
Nucl.\ Phys.\ A {\bf 696} (2001) 788.

\bibitem{Grau:2005sm}
  N.~Grau  [PHENIX Collaboration],
  nucl-ex/0511046.


\bibitem{Casalderrey-Solana:2006sq}
  J.~Casalderrey-Solana, E.~V.~Shuryak and D.~Teaney,
  hep-ph/0602183.

\bibitem{Ollitrault:1992bk}
  J.~Y.~Ollitrault,
  Phys.\ Rev.\ D {\bf 46} (1992) 229.

\bibitem{Eskola:2005ue}
 See e.g. K.~J.~Eskola, {\it et al}
  Phys.\ Rev.\ C {\bf 72} (2005) 044904.

\bibitem{Shuryak:2005pp}
  See e.g. E.~Shuryak,
  hep-ph/0510123.

\bibitem{Maiani:2006ia}
  L.~Maiani, A.~D.~Polosa, V.~Riquer and C.~A.~Salgado,
  hep-ph/0606217.

\bibitem{Friesvb}
  R.~J.~Fries, {\it et al.}
  Phys.\ Rev.\ C {\bf 68}, 044902 (2003).

\bibitem{Salur:2005nj}
  S.~Salur  [STAR Collaboration],
  nucl-ex/0509036.

\bibitem{Policastro:2001yc}
  G.~Policastro, D.T.~Son and A.O.~Starinets,
  Phys.\ Rev.\ Lett.\  {\bf 87} (2001) 081601.

\bibitem{Casalderrey-Solana:2006rq}
  J.~Casalderrey-Solana and D.~Teaney,
  arXiv:hep-ph/0605199.

\bibitem{Liu:2006ug}
  H.~Liu, K.~Rajagopal and U.~A.~Wiedemann,
  hep-ph/0605178;
  N.~Armesto, J.~D.~Edelstein and J.~Mas,
  arXiv:hep-ph/0606245.


\bibitem{Friess:2006fk}
  J.~J.~Friess, S.~S.~Gubser, G.~Michalogiorgakis and S.~S.~Pufu,
  arXiv:hep-th/0607022.

\bibitem{Janik:2005zt}
  R.~A.~Janik and R.~Peschanski,
  Phys.\ Rev.\ D {\bf 73} (2006) 045013.





\end{thebibliography}
\end{document}